\documentstyle[aps,epsfig]{revtex}

\begin{document}

\draft

\twocolumn[\hsize\textwidth\columnwidth\hsize\csname@twocolumnfalse%
\endcsname
\title{Soap froths and crystal structures}

\author{P. Ziherl\footnote{On leave from J. Stefan Institute, Slovenia.}
and Randall D. Kamien}

\address{Department of Physics and Astronomy, University of Pennsylvania,
Philadelphia, PA 19104-6396, USA}

\date{\today}

\maketitle

\begin{abstract}
We propose a physical mechanism to explain the crystal symmetries found in
macromolecular and supramolecular micellar materials. We argue that the
packing entropy of the hard micellar cores is frustrated by the entropic
interaction of their brush-like coronas. The latter interaction is treated
as a surface effect between neighboring Voronoi cells. The observed crystal
structures correspond to the Kelvin and Weaire-Phelan minimal foams. We show
that these structures are stable for reasonable areal entropy densities.
\end{abstract}
]

Dendritic polymers~\cite{Balagurusamy97,Percec98,Hudson99}, hyper-branched
star polymers~\cite{Dotera99,Watzlawek99} and diblock
copolymers~\cite{McConnell93,McConnell96} represent a new class of
molecular assemblies all of which form a variety of crystalline lattices,
many of which are not close-packed. These assemblies are all characterized
by compact cores and brush-like, soft coronas. These systems might be
modeled by treating the micelles as sterically interacting hard spheres
and it would follow that their crystalline phases should be stackings of
hexagonal-close-packed (HCP) layers. Recently~\cite{Mau99} it has been
shown that the face-centered cubic (FCC) lattice maximizes the total
entropy and so hard-sphere crystals should form FCC structures. Note that
the entropic difference between the various HCP lattices is a {\sl global}
issue: the local arrangement of spheres is the same for all close-packed
variants and thus the lattice cannot be predicted from nearest-neighbor
interactions. In order to understand the richness of crystal symmetries in
the micellar systems, we propose an additional global consideration: we
add an interaction proportional to the interfacial surface area between
the cages which contain each micelle (Voronoi cells). Though approaches
based on self-consistent field theory and two-body interactions can yield
non-close-packed lattices~\cite{Matsen97,Velasco00}, we propose a
universal explanation for a host of new structures and present a new
paradigm for the rational design and control of macromolecular
assemblies~\cite{Muthukumar97}.

The interfacial interaction arises through the entropy of the brush-like
coronas of the micelles. Because of constraints on their conformations,
the brushes suffer an entropic penalty proportional to the interfacial
area between the Voronoi cells surrounding each sphere. Thus they favor
area-minimizing structures, precisely the type of structures that dry
foams might make. Over a century ago, Lord Kelvin proposed that a
body-centered-cubic (BCC) foam structure had the smallest
surface-to-volume ratio~\cite{Thomson87} but in 1994 Weaire and Phelan
found that a structure based on the A15 lattice~\cite{Weaire94a} was more
efficient. We note that neither the BCC nor A15 structures are
close-packed and thus there is a fundamental frustration between the
hard-core volume interaction and the surface interaction due to
overlapping soft coronas.

\begin{figure}

\bigskip
\bigskip

\centerline{\epsfig{file=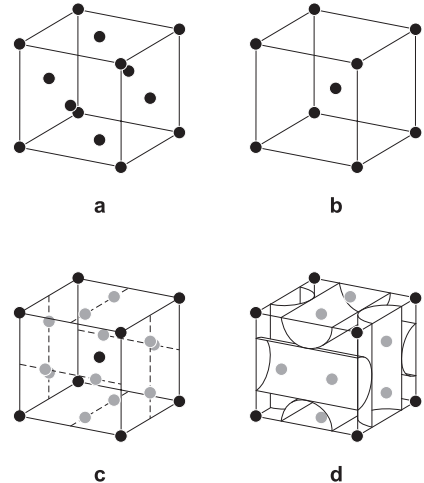,width=55mm}}

\vspace*{5mm}

\caption{Various lattices: a) Face-centered cubic, b) body-centered cubic,
c) A15 lattice, and d) columnar representation of A15 lattice. In the
A15 lattice, columnar and interstitial sites are drawn in grey and black,
respectively.}
\label{A15}
\end{figure}

For concreteness, in this paper we focus on structures observed in a
family of dendrimer compounds consisting of a compact poly(benzyl ether)
core segment and a diffuse dodecyl corona~\cite{Balagurusamy97,Percec98}.
These conical dendrimers self-assemble in spherical micelles which
subsequently arrange into the A15 lattice (Fig.~\ref{A15}). The
interaction between the micelles is primarily steric, {\sl i.e.},
repulsive and short-range. The micellar architecture suggests that the
potential is characterized by three regimes. At large distances, the
micelles do not overlap and the interaction vanishes. As the coronas begin
to overlap, the entropy of the brush-like coronas decreases, which gives
rise to a soft repulsion between the micelles. Finally, at small
separations the coronas begin to penetrate the compact cores: this is very
unfavorable and gives rise to hard-core repulsion. This energy landscape
is in qualitative agreement with recent, detailed molecular dynamics
simulations~\cite{Cagin00}.

Although both originate in steric interaction, the two repulsive regimes
are characterized by very different functional behaviors. The hard part of
the potential results in a restricted positional entropy of the micelles
which depends on the free volume, the difference between the actual and
the hard-core volumes. The soft part comes from the decreased
orientational entropy of the chains within the overlapping coronas. The
matrix of overlapping coronas can be thought of as a compressed bilayer
and thus the free volume may be written as a product of the interfacial
area $A$ and the average spacing between the hard cores $d$ so that at any
given density
\begin{equation}
Ad={\rm constant.}
\end{equation}
Though this approximation ignores the curvature of brush-like coronas, the
dendrimers are relatively close and we expect this constraint to hold in
this system. Since the repulsion decreases monotonically with distance,
the system will favor a maximum thickness $d$ and will thus tend to
minimize the interfacial area. Hence our proposed interfacial interaction,
which is incompatible with the bulk free energy minimized by a
close-packed arrangement of micelles. In the following, we compare the
free energies of FCC, BCC, and A15 lattices and estimate the strength of
the interfacial interaction such that the structure of the micellar
crystal is dictated by the minimal-area principle.

The calculation of the bulk free energies of condensed systems is fairly
complicated even for hard-sphere systems and the best theoretical results
are obtained numerically. It is interesting to note that elaborate
analytic models, such as the high-density analog of the virial
expansion~\cite{Rudd68} and the weighted-density-functional
approximation~\cite{Curtin87}, are only slightly better than the simple
cellular free-volume theory~\cite{Curtin87,Barker63}. The free-volume
theory is a high-density approximation where each micelle is contained in
a cell formed by its neighbors, and the communal entropy associated with
the correlated motion of micelles is neglected.

Within this theory, the positional entropy of a micelle is determined by
the configurational space of its Voronoi or Wigner-Seitz cell. In the FCC
lattice, the centers of mass of the micelles are within rhombic
dodecahedra~\cite{Kittel53}, while in the BCC lattice they are contained
in regular octahedra although the BCC Voronoi cell is an orthic
tetrakaidecahedron~\cite{Kittel53}. For these lattices, the bulk free
energy of a micelle is given by
\begin{equation}
F^{\scriptscriptstyle X}_{\rm bulk}=-k_{\scriptscriptstyle B}T\ln\Bigglb(
\alpha^{\scriptscriptstyle X}\left({\beta^{\scriptscriptstyle X}\over
n^{1/3}}-1\right)^3\Biggrb),
\label{fccbcc}
\end{equation}
where $X$ is either FCC or BCC, $n=\rho R^3$ is the reduced number
density, and $R$ is the hard-core radius of micelles. The coefficients
$\alpha^{\rm FCC}=2^{5/2}$ and $\alpha^{\rm BCC}=2^23^{1/2}$ depend on the
shape of the cells, whereas $\beta^{\rm FCC}=2^{-5/6}$ and $\beta^{\rm
BCC}=2^{-5/3} 3^{1/2}$ are determined by their size.

\begin{figure}

\bigskip
\bigskip

\centerline{\epsfig{file=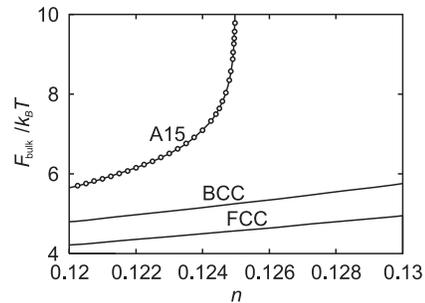,width=55mm}}

\vspace*{5mm}

\caption{Bulk free energies of FCC, BCC, and A15 lattice at reduced
densities above the melting point. Solid lines: analytical results,
Eqs.~(\ref{fccbcc}) and (\ref{a15}); circles: numerical results.}
\label{fbulk}
\end{figure}

The A15 lattice is somewhat more complicated: as shown in Fig.~\ref{A15}d
the A15 unit cell includes 6 columnar sites, which make up 3 perpendicular
interlocking columns, and 2 interstitial sites. A pseudo-Voronoi
construction (subject to the constraint that all cells have equal volume)
for this lattice leads to a partition consisting of irregular pentagonal
dodecahedra and tetrakaidecahedra with two hexagonal and twelve pentagonal
faces~\cite{Charvolin88}. Because of the irregularity of the cells, we
calculate the bulk entropic free energy numerically, and the result is
shown in Fig.~\ref{fbulk}. For our purposes we require an analytic form:
by substituting the dodecahedra and tetrakaiecahedra by spheres and
cylinders, respectively, and allowing for two adjustable parameters $C$
and $S$, which measure the deviation of the Voronoi cells from these
spheres and cylinders, we find
\begin{eqnarray}
F_{\rm bulk}^{\rm A15}&&=-k_{\scriptscriptstyle B}T\left[\frac{1}{4}\ln
\Bigglb(\frac{4\pi S}{3}\left(\frac{\sqrt{5}}{4n^{1/3}}-1\right)^3\Biggrb)
\right.\nonumber\\
&&+\left.\frac{3}{4}\ln\Bigglb(2\pi C\left(\frac{\sqrt{5}}{4n^{1/3}}-1
\right)^2\left(\frac{1}{2n^{1/3}}-1\right)\Biggrb)\right].
\label{a15}
\end{eqnarray}
This form is within $0.1\%$ of the numerical result with $S\approx1.64$
and $C\approx1.38$.

The interfacial free energy is minimized by the division of space with
smallest area. The problem of finding the partition of space into
equal-volume cells with the minimum interfacial area was first studied by
Kelvin~\cite{Weaire97}: he proposed a BCC lattice of orthic
tetrakaidecahedra with slightly curved hexagonal faces to satisfy the
Plateau rules~\cite{Thomson87}. However, the Weaire-Phelan partition,
which differs from the equal-volume Voronoi construction for the A15
lattice only in a delicate curvature of the pentagonal faces, is 0.3\%
more efficient~\cite{Weaire94a}. We note that the BCC and A15 structures
are among the simplest tetragonal close-packed lattices~\cite{Rivier94},
suggesting that other, more complex close-packed lattices might be more
efficient still. However, the A15 structure appears to be the most
efficient, although no proof of its supremacy exists.

To argue that the A15 and BCC lattices are the equilibrium structures
formed by micelles, we must estimate the entropy penalty per unit area and
translate this into an entropy per dodecyl chain. The dodecyl bilayer is
modeled as a polymer brush consisting of chain molecules attached to hard
cores, and in the limit of high interdigitation its free energy consists
solely of the excluded-volume repulsion of the chains:
\begin{equation}
F_{\rm surf}=\frac{2\ell Nk_{\scriptscriptstyle B}T}{d},
\end{equation}
where $d$ is the layer thickness, $\ell$ is a parameter with the dimension
of length and $N$ is the number of chains per micelle~\cite{Milner88}.
Since the bilayer must fill the free volume, $A_Md=2(n^{-1}-4\pi/3)R^3$,
where $A_M$ is the interfacial area per micelle. Thus the interfacial free
energy of a micelle is
\begin{equation}
F^{\scriptscriptstyle X}_{\rm surf}=\frac{\ell Nk_{\scriptscriptstyle B}T}
{R}\frac{\gamma^{\scriptscriptstyle X}n^{-2/3}}{n^{-1}-4\pi/3},
\end{equation}
where $\gamma^{\rm FCC}=2^{5/6}3=5.345$, $\gamma^{\rm BCC}=5.306$, and
$\gamma^{\rm A15}=5.288$~\cite{note}.

We now calculate the range of $\ell$ such that the total free energy
\begin{equation}
F^{\scriptscriptstyle X}=F_{\rm bulk}^{\scriptscriptstyle X}+F_{\rm
surf}^{\scriptscriptstyle X}
\end{equation}
is minimized by the BCC and A15 lattices rather than by the na\"\i ve,
close-packed, FCC lattice. In order to estimate the strength of the soft
repulsion, we must first determine the actual reduced density $n$. Since
the hard-core radius of the micelles is unknown, we limit $n$ by
recognizing that it must be larger than the melting density, $n\approx
0.120$ for hard spheres~\cite{Curtin87}, and that it must be smaller than
the close-packing density of the A15 lattice, $n=0.125$. The most
conservative lower bound of $\ell$ corresponds to the lowest possible
density, {\sl i.e.}, the melting density. With $N=162$ chains per
micelle~\cite{Balagurusamy97}, we find that at $n=0.120$ the FCC to BCC
transition occurs for $\ell\approx0.1R$ and the BCC to A15 transition
occurs for $\ell\approx0.3R$. This corresponds to an entropy per chain of
$0.5k_{\scriptscriptstyle B}$ and $1.5k_{\scriptscriptstyle B}$,
respectively. Both values are of the correct order of magnitude and the
higher value of the latter is consistent with the relative rarity of the
A15 phase.

Since we expect that each chain has at least $k_{\scriptscriptstyle B}$ of
entropy, we conclude that the energetics of the dendrimer micelles is
dominated by interfacial effects. This is hardly surprising. The number of
degrees of freedom of each micelle is quite large and the bulk free energy
only depends on the position of the micelle as a whole. Since the micelles
are soft, the internal degrees of freedom such as the chain conformations
play an important role.

This paradigm -- which shows that the minimal surface problem can be
fruitfully transplanted to the microscopic level -- explains the
morphology of a number of dense micellar systems. The same ideas can be
applied to polymeric micelles made of, {\sl e.g.},
polystyrene-polyisoprene diblock copolymers dispersed in
decane~\cite{McConnell93,McConnell96}. In this case, the micelles are
characterized by highly concentrated polystyrene core and diffuse
polyisoprene corona, and they form BCC or FCC lattices, depending on the
relative length of the polystyrene and polyisoprene chains. The BCC
lattice is observed in diblock copolymers with similar lengths of core and
coronal segments, whereas the FCC lattice occurs whenever the corona is
thin compared to core. This is consistent with our model of the
impenetrable core which is responsible for the hard part of the repulsion
and which favors arrangements with large free volume. In addition, our
model suggests that the A15 lattice is the ground state of an asymmetric
diblock with an exceptionally large corona or, equivalently, a corona made
of very floppy, ``entropy-rich'' chains. We note that distinguishing
between A15 and BCC in powder-averaged diffraction is delicate: the first
three BCC reflections~\cite{Hahn83} are at $\sqrt{2}$, $\sqrt{4}$, and
$\sqrt{6}$, while the first {\sl four} A15 reflections are at $\sqrt{2}$,
$\sqrt{4}$, $\sqrt{5}$, and $\sqrt{6}$, and thus a careful study would be
necessary.

The existence of the A15 lattice in the dendrimer aggregate also may be
regarded as an experimental verification of the recent theoretical
developments in minimal surfaces and, in particular, Weaire and Phelan's
conjecture that this structure solves the Kelvin problem. At this
juncture, the presumably ideal A15 structure has not been observed
unambiguously on a macroscopic scale in a soap froth~\cite{Weaire94b}.
Last but not least, let us note that similar structures have been found in
lyotropic materials, {\sl e.g.}, in lipid bilayers in
water~\cite{Mariani94}. In such systems the intermicellar potential also
results in an effective interfacial free energy although it is not steric
but substance-specific, and thus transcends the scope of this discussion.

Our model may be further refined by including the effects of the curvature
of the brush-like coats, the strain of the coronas into the interstitial
regions, and solvent effects. In addition, the dual problem of determining
the structure of foams might be amenable to our analysis through the
introduction of excluded volume interactions between the
bubbles~\cite{Durian95}. Recent work focusing on two-body interactions has
shown that

We hope that this study elucidates the relation
between interaction and structure in supramolecular assemblies. By including
an additional global contribution to the free energy we provide a rough yet
universal quantitative guideline for the design of self-organized soft
materials, which can be used for a number of applications such as photonic
band-gap materials~\cite{Tarhan96}, Bragg switches~\cite{Lin00}, and porous
microreactors~\cite{Jenekhe99}. By tuning the ranges of hard and soft
repulsion, one should be able to choose among the spectrum of symmetries
from the lattice with minimal interfacial area to the lattice with maximal
packing fraction and engineer the crystal structure most fitted for a
particular application.

It is a pleasure to acknowledge stimulating
conversations with T.~C.~Lubensky and V.~Percec. This work was supported in
part by NSF Career Grant DMR97-32963. RDK was also supported by the
Alfred~P.~Sloan Foundation.


\begin{references}
\bibitem{note} $\gamma^{\rm BCC}$ and $\gamma^{\rm A15}$ were computed using
Surface Evolver, a software package developed by K. Brakke.
\end{references}
\end{document}